\documentclass[aps,prb,twocolumn,amsmath,amssymb,showpacs]{revtex4}
\usepackage{graphicx}
\usepackage{dcolumn}
\usepackage{bm}
\begin{document}

\title{A quantitative evaluation of metallic conduction in conjugated polymers}

\author{H.C.F. Martens}

\affiliation{Philips Research Laboratories, Prof. Holstlaan 4, 5656 AA Eindhoven, The Netherlands}

\author{H.B. Brom}

\affiliation{Kamerlingh Onnes Laboratory, Leiden University, PO Box 9504, 2300 RA Leiden, The Netherlands}

\date{August 23 2004, will appear as Rapid Comm. in Phys. Rev. B}

\begin{abstract}
As the periodicity in crystalline materials creates the optimal
condition for electronic delocalization, one might expect that in
partially crystalline conjugated polymers delocalization is
impeded by intergrain transport. However, for the best conducting
polymers this presumption fails. Delocalization is obstructed by
interchain rather than intergrain charge transfer and we propose a
model of weakly coupled disordered chains to describe the physics
near the metal-insulator transition. Our quantitative calculations
match the outcome of recent broad-band optical experiments and
provide a consistent explanation of metallic conduction in
polymers.
\end{abstract}


\maketitle

\begin{figure}
\begin{center}
\includegraphics[width=7.5cm]{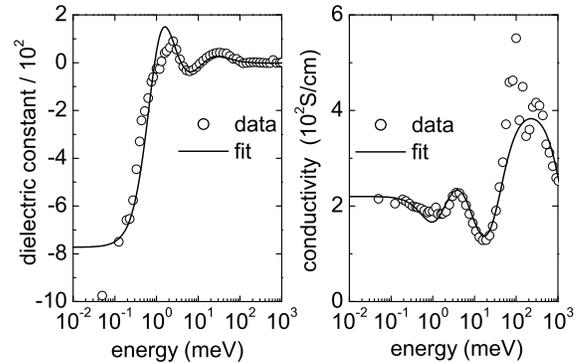}
\end{center}
\caption{Optical dielectric function and conductivity of  metallic
polypyrrole. Symbols: experimental results taken from
Ref.~\onlinecite{Romijn03}. Drawn lines: theoretical prediction
for weakly coupled disordered chains (see text).
\label{optical_conductivity}}
\end{figure}

\begin{figure}
\begin{center}
\includegraphics[width=7.5cm]{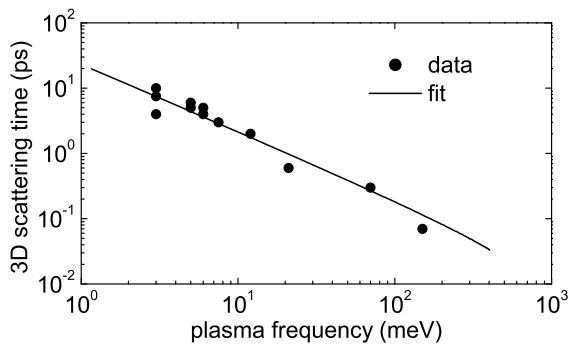}
\end{center}
\caption{Scaling relation between plasma frequency and 3D
scattering time. Dots: experimental results taken from
Ref.~\onlinecite{Martens01}. Full line: calculation for disorder
driven MIT (see text). \label{scaling}}
\end{figure}

One of the hallmarks of metallic transport is the negative
dielectric function $\varepsilon$ at low frequency $\omega$, that
reflects a time lag between induced current and applied field due
to the inertia of delocalized charge carriers. The surprising
observation of negative $\varepsilon$ in conjugated polymers by
Kohlman and co-workers, \cite{Kohlman97} which was recently
verified by Romijn {\it et al.} and Lee and Heeger,
\cite{Romijn03,Lee03} spurred experimental and theoretical
research into the nature of metallic conduction in these
materials. As displayed in Fig.~\ref{optical_conductivity}, the
full spectral response is quite complicated including multiple
zero crossings of $\varepsilon(\omega)$ and non-monotonous
behavior of the conductivity $\sigma(\omega)$. The low-$\omega$
dynamics are characterized by long scattering times $\tau\sim$~ps
and low plasma frequencies $\omega_{\rm p}\sim$~meV. This differs
three orders of magnitude from conventional metals where
$\tau\sim$~fs and $\omega_{\rm p}\sim$~eV, while the carrier
density $n$ is only one order of magnitude less. An empirical
scaling relation between $\tau$ and $\omega_{\rm p}$ has been
found, see Fig.~\ref{scaling}, suggesting a common mechanism
governs these parameters. \cite{Martens01} \\
The initial models for metallic polymers relied either on the
Anderson theory of localization, \cite{Lee95,Handbook98} or on a
granular picture, \cite{Kohlman97,Handbook98} but both predict
$\varepsilon>0$. \cite{Lee95,Tzamalis02,Chapman99,Levy97}
Including percolation effects explains negative $\varepsilon$,
\cite{Kohlman97,Levy97} but the calculated $\omega_{\rm p}$ lies
two orders of magnitude above the experimental values.
\cite{Levy97} Prigodin and Epstein suggested that the metallic
state is sustained by atypical resonant tunnelling events and
used the Landauer-B\"uttiker transmission framework to explain
small $\omega_{\rm p}$ and long $\tau$.\cite{Prigodin02} However, the
conditions for resonant tunnelling are not always
fulfilled, \cite{Romijn03} and the detailed
physical mechanisms remain to be untangled.\\
Recently, Prigodin and co-workers successfully explained the
$\omega$-dependence in the insulating phase using quasi-1D
variable-range hopping theory with interchain transfer as
rate-limiting step. \cite{Prigodin04}
By accounting for quasi-1D conduction, Kaiser {\it et al.} \cite{Kaiser01}
gave a description for metallic conductivity of granular polymers, which works for
blends as well. \cite{Long04} Recent optical experiments by Lee and
Heeger, \cite{Lee03} and us, \cite{Martens01} could be explained
in a 1D picture as well. Also Kohlman {\it et al.}
discussed 1D effects. \cite{Kohlman97} Here, we attempt to
elucidate the nature of metallic conduction in polymers in more
detail. We first argue why the metallic phase is dominated by the 1D
structure. Then, like Ref.~\onlinecite{Prigodin02}, we use a
Landauer-B\"uttiker approach to model the transport and show how
interchain coupling drives the intrinsically insulating 1D phase to a
metallic state at the cross-over from 1D to 3D coherent conduction.
All transport parameters can be described in terms of disorder and coupling
strength and the relevant length scales are $\zeta$ the 1D
localization length, and $\lambda$ the length over which carriers
experience sequential interchain events. For weak coupling
$\zeta<\lambda$, transport keeps a 1D signature and effectively
only a fraction of the carriers is involved in 3D transport. Our
model reproduces the reported experimental results well and naturally
ascribes the unusual low-$\omega$ dynamics to small interchain
coupling.

In metallic conductors, delocalized states exist at the Fermi
level. Delocalization is counteracted by disorder and the balance
between metallic and insulating groundstates
is critically dependent on the weakest charge transfer steps in
the system. In the disordered polymers, chains are aligned within
microcrystallites that are separated by amorphous regions, and
delocalization may be impeded by either {\it intergrain} charge
transfer or by {\it interchain} charge transfer. To identify the
critical mechanism, we compare the energy scales.

{\it Intergrain charge transfer --} For the best conducting
polymers the crystalline fraction is $f_c = 50 \% $ (polypyrrole,
PPy, and polyaniline, PAni), even up to $f_c = 90 \% $ for
polyacetylene (PAc); the typcial microcrystallite size $D_c$
amounts to 2 nm (PPy), 5 nm (PAni), or 10 nm (PAc). For isolated
grains, \cite{Prigodin02} geometrical considerations imply a
width of amorphous regions $W_a = D_c(f_c^{-1/3} -
1)$, giving $W_a$ = 0.5~nm, 1.3~nm, and~0.4 nm, respectively.
Prigodin and Epstein estimate the localization length along the
chain $\zeta = 1.2$~nm. \cite{Prigodin02} A different approach is
based on the potential barrier height $B$ of the amorphous region,
which from transport activation energy measurements we estimate $B
= 0.01$--$0.1$~eV. \cite{Martens0102} A wavefunction penetrating
this barrier falls off exponentially with $\zeta = (2m_{\rm
e}B/\hbar^2)^{-1/2} = 0.6$--$1.9$~nm. This
corresponds to fully back-scattered waves, expected for strong
disorder, and agrees with the experimental value $\zeta = 0.8$~nm
for insulating polythiophene. \cite{Reedijk99} Another approach is
based on a result by Thouless: $\zeta = a(E_F/B)^2$,
\cite{Thouless73} which easily yields 10~nm using reasonable
parameters ($E_F=0.5$~eV, $B=0.1$~eV, and $a=4$~\AA\ the
microscopic length scale, e.g. a monomer). Taking a conservative
estimate $\zeta = 1.2$~nm, the intergrain transfer integral
through amorphous regions along 1D chains $I_\parallel =
I_0\exp(-2W_a/\zeta)$, with $I_0\approx2.5$~eV the intra-chain
$\pi$-electron transfer of the unperturbed chain, equals at least
1.1~eV, 0.3~eV, and 1.3~eV for PPy, PAni, and PAc respectively.

{\it Interchain charge transfer --} In conjugated polymers, the
$\pi$-electron system is formed by hybridization of $2p_x$
orbitals on adjacent carbon atoms that constitute the polymer
backbone. Interchain charge transfer stems from the overlap of the
$\pi$-electron clouds on neighboring chains. We estimated the
interchain charge transfer $I_\perp$ for parallel chains of carbon
$2p_x$ orbitals using the Slater wavefunction approximation. Our
calculations suggest an optimum packing distance of 3 \AA\ where
$I_\perp\approx10$~meV is maximized. For larger separation,
$I_\perp$ falls off exponentially with interchain localization
length $\sim 1$~\AA\ in agreement with experimental results.
\cite{Reedijk99} Extensive tight-binding calculations by Mizes and
Conwell yielded $I_\perp\approx$~30~meV for PAc. \cite{Mizes91}

Thus, for $f_c\approx50$~\% the energy scale for
interchain charge transfer is at least an order of magnitude
smaller than that for intergrain charge transfer. \cite{note}
This is not surprising when it is realized that
intergrain transport is essentially an {\it intra}chain process
that proceeds along the covalently bound chains, while adjacent
chains are not chemically but Van der Waals-bonded and hence {\it
inter}chain electronic coupling should be much weaker. It means
that $(i)$ intrachain carrier delocalization extends over several
grains and $(ii)$ formation of truly delocalized states is
governed by the small interchain charge transfer. Therefor, we
propose a system of weakly coupled disordered chains to model
conducting polymers. Our approach is as follows: we limit
ourselves to an isolated disordered 1D chain where quantum
mechanics leads to localized 1D wavefunctions $\Psi_{\rm 1D}$. We
argue that, staying within a 1D framework, interchain carrier
exchange is equivalent to dephasing of the $\Psi_{\rm 1D}$s and
derive the condition for delocalization.
Finally, we apply our model to describe the carrier dynamics of
conducting polymers using known microscopic parameters and a
tight-binding approximation for the interchain coupling.

First, we briefly review the case for isolated chains. Even for minimal
disorder, the electronic eigenstates $\Psi_{\rm 1D}$ are localized
due to repeated coherent back-scattering. \cite{Thouless73} A
convenient approach to solve the transport problem is the
Landauer-B\"uttiker transmission framework. \cite{Buttiker88}
Assume a transmission coefficient $T \leq 1$ per unit length $a$.
Then, over a distance $z>a$ the
envelope of an unreflected wave falls off as $T^{z/a}$ implying an
intrachain localization length $\zeta=a/\ln(1/T)$. Disorder
determines the microscopic transmission probabilities and hence
sets $\zeta$. The mean level spacing of the localized states is
$\Delta = 1/(g\zeta)$ with $g$ the density of states per unit
length and unit energy. The dimensionless conductance $G=T/(1-T)$
and it follows that at short length scales ($z<\zeta$) $G$ is
finite and given by the classical (Ohmic) result $G(z)=\zeta/z$,
while for a long chain $G(z)=\exp(-z/\zeta)$ scaling is non-Ohmic
and the conductivity, $G\, z$, becomes zero for
$z\rightarrow\infty$. The conductivity due to resonant
transitions between localized states is appreciable only for
$\hbar\omega\sim\Delta$. For $\hbar\omega>\Delta$, this Lorentz
oscillator response approaches the Drude
conductivity with plasma frequency $\Omega_{\rm p}\sim$~eV and
intrachain scattering time $\tau_0\sim$~fs.

The above discussion is essentially based on a one-electron
Schr\"odinger equation assuming phase-coherent transport along the
chain. Interchain transfer introduces a probability $\epsilon$ for
interchain transitions and formally the isolated chain
eigenfunctions are no longer appropriate. For small $\epsilon$,
however, the isolated chain picture remains accurate for short
time and length scales: the $\Psi_{\rm 1D}$s obtain a finite
lifetime. In case of strong coupling such a picture is not
realistic and the carrier transport problem should be considered
for the fully coherent 3D system. In conjugated polymers
interchain charge transfer is weak and a 1D transport model
including the effect of coupling seems a good starting point.
Particle conservation implies that interchain coupling occurs in
the form of carrier exchange with a reservoir, where the carrier's
initial state $\Psi_{\rm 1D,1}$ differs from the final state
$\Psi_{\rm 1D,2}$, which thus introduces the finite lifetime of
the 1D states: interchain transport leads to dephasing reflecting
that the $\Psi_{\rm 1D}$s are not ``true eigenstates'' of a system
of coupled chains. Usually, phase-breaking is associated with
inelastic events occurring at non-zero temperature. However,
phase-breaking is not necessarily dissipative when the initial and
final states have the same energy. \cite{Datta90} Allowing
``elastic phase-breaking'' implies that the ``true eigenstates''
are extended over the entire system of coupled chains since then
the energy level spacing $b^2/(gL^3)$ vanishes for
$L\rightarrow\infty$ ($b$ interchain distance), i.e. the system is
gapless, or in other words a metal. The condition for which this
becomes possible determines the zero-temperature MIT and will be
discussed further below.

Phase-breaking events are modelled in the Landauer-B\"uttiker
framework by incorporating current-conserving phase-randomizing
scatterers into the chain: each scatterer randomizes phase with
amplitude $\sqrt{\epsilon}$ yielding probability $\epsilon$ for
phase-breaking. The total transmission probability is given by
$T_{\rm tot} = T_{\rm c} + T_{\rm i}$, \cite{Buttiker88} with
$T_{\rm c}$ the direct coherent intrachain transmission
probability (no interchain events) and $T_{\rm i}$ transmission
including at least one interchain transition. The conductance
\begin{equation}
G = \frac{T_{\rm tot}}{1-T_{\rm tot}}=\frac{T_{\rm c}}{1 - T_{\rm
c} - T_{\rm i}} + \frac{T_{\rm i}}{1 - T_{\rm c} - T_{\rm i}}
\label{LandauerFormula2}
\end{equation}
consists of two contributions: proportional to $T_{\rm c}$ and
$T_{\rm i}$, respectively. However, the relations are not
straightforward, as the denominators contain both $T_{\rm c}$ and
$T_{\rm i}$.

The carrier fraction $\epsilon$ that suffered dephasing at $z=0$
will, on average, suffer a subsequent dephasing event at
$z\sim\lambda=a/\epsilon$. We can discern two
limits: $z<\lambda$ and $z>\lambda$. When $z<\lambda$, $T_{\rm c}
\gg T_{\rm i}\sim\cal{O}(\epsilon)$,
Eq.~(\ref{LandauerFormula2}) simplifies to that of the isolated chain: $G = T_{\rm
c}/(1-T_{\rm c})$. When
$z>\lambda$, $T_{\rm c}$ is exponentially small and $G = T_{\rm i}/(1-T_{\rm i})$. Strong
coupling, i.e. every scatterer dephases ($\epsilon=1$),
$\lambda=\zeta$, corresponds to the classical (incoherent) case:
$G=\lambda/z=\zeta/z$. For strong disorder and weak coupling,
$\lambda>\zeta$, the transmission amplitude $T_{\rm i}$ becomes
suppressed by localization effects. Then, $T_{\rm i}$
consists of the product of the dephasing probability $\epsilon$
and a factor $\zeta/a$ that counts the number of channels
available for the interchain transport on the scale $\zeta$ where
the wavefunctions have appreciable amplitude,
multiplied by $\exp(-\lambda/\zeta)$ the 1D coherent transmission
amplitude between subsequent dephasing events, yielding
$G=\epsilon(\zeta/a)(\lambda/z)\exp(-\lambda/\zeta)=(\zeta/z)\exp(-\lambda/\zeta)$.
Thus, for weak coupling, the classical expectation value for the
conductance is suppressed by 1D localization effects. Delocalized
carrier transport remains piecewise 1D and the length scale of interest
is $\lambda$ with a 1D transport time
$\tau_\lambda=(\zeta/v_{\rm F})\exp(\lambda/\zeta)$. At $\tau_\lambda$
scale, only fraction $\epsilon$ of the carriers is involved in interchain
transport and thus crossover to full 3D conduction occurs at $\tau_{\rm
3D}=\tau_\lambda/\epsilon$.

\begin{figure}
\begin{center}
\includegraphics[width=7.5cm]{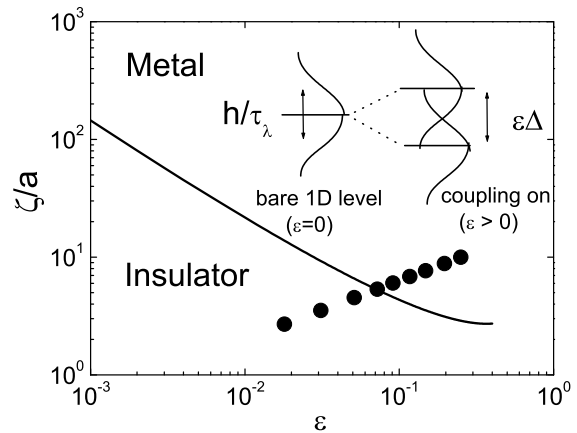}
\end{center}
\caption{Metal-insulator phase-diagram of coupled 1D
chains as a function of interchain coupling and
intrachain localization length. Below the drawn line the system
is insulating, above it is metallic. Dots:
disorder driven MIT in conducting polymers. The inset gives a
graphical representation of the leitmotif behind the MIT criterion
discussed in the text. \label{MIT}}
\end{figure}

The above results demonstrate that in the presence of dephasing,
the dc conductivity remains finite for $z\rightarrow\infty$. This
holds for any dephasing mechanism; in particular inelastic events
due to finite temperature. \cite{Maschke94} As argued above,
``elastic dephasing'' is a necessary condition for the formation
of a truly metallic state. In our 1D model, the interchain
transfer couples a $\Psi_{\rm 1D,1}$ to a different $\Psi_{\rm
1D,2}$ on lengthscale $\lambda$ and this gives a typical level
splitting $\epsilon\Delta$ of these states. When the finite width
of the 1D energy levels (Thouless energy) on length scale
$\lambda$, given by $h/\tau_\lambda$, exceeds this level-splitting
(bandwidth) energy-matched (elastic) interchain transitions are
possible and hence a metallic phase can be formed, as indicated in
the inset of Fig.~\ref{MIT}. Applying the relation $hv_{\rm
F}=g^{-1}$ valid for 1D conductors, the criterion for
delocalization translates into $a/\zeta=(B/E_{\rm
F})^2<\epsilon\ln(1/\epsilon)$. Alternatively, one can use the
condition $G>1$ for a metal on a characteristic length scale. When
applied to the carrier fraction that is effectively involved in
the interchain transport on length scale $\lambda$ this gives
$G_\epsilon(\lambda) = G(\lambda)/\epsilon =
\exp(-\lambda/\zeta)/\epsilon> 1$, leading to the same result. The
phase diagram for coupled disordered chains as function of
disorder ($\zeta/a$) and interchain coupling ($\epsilon$) is shown
in Fig.~\ref{MIT}.

Let us now apply our model to the metallic polymers. First we
derive $\epsilon$ using a tight-binding approximation. Interchain
charge transfer would, in the absence of disorder, lead to an
interchain bandwidth $\propto I_\perp$. However, as a result of
disorder electronic states $\Psi_{\rm 1D}$ on adjacent chains
typically have an energy mismatch $\approx\Delta/2$. Provided that
$I_\perp<\Delta/2$, the bandwidth available for interchain
transport reduces to $4I_\perp^2/\Delta$ which corresponds to an
interchain coupling $\epsilon = 4(I_\perp/\Delta)^2$ governed by
both interchain transfer and disorder. Next, we derive the
relevant transport time (energy) scales. Crossover to full 3D
conduction occurs on time scale $\tau_{\rm
3D}=\tau_\lambda/\epsilon$. Following the reasoning for the MIT
criterion, the bandwidth for interchain transport is
$\epsilon\Delta\exp(-\lambda/\zeta)$ and, using tight-binding, the
effective mass for 3D transport becomes
$m^*=8\hbar^2\exp(\lambda/\zeta)/(\epsilon\Delta b^2)$. 1D
localization suppresses the transmission probability on scale
$\lambda$ and the effective carrier density involved in 3D
transport reduces to $n_{\rm eff}=n\exp(-\lambda/\zeta)$. This gives:
\begin{eqnarray}
\tau_{\rm 3D}  & = & (\pi\hbar/2I_\perp^2g\zeta)\exp\left(a/4I_\perp^2g^2\zeta^3\right),\label{tau3D}\\
\omega_{\rm p} & = & \sqrt{ne^2b^2I_\perp^2g\zeta/2\hbar^2\varepsilon_0}\label{omegap}
\exp\left(-a/4I_\perp^2g^2\zeta^3\right).
\end{eqnarray}

For $\omega\tau_{\rm 3D}<1$, the 3D transport channel is open
and the conductivity is $\sigma(\omega)=
\sigma_{\rm 3D}/(1+i\omega\tau_{\rm 3D})$. At intermediate
energy, $\hbar/\tau_{\rm 3D} < \hbar\omega < \Delta$, the 1D
signature of delocalized carrier transport becomes visible.
Screening in 1D conductors is poor and the dynamic response
is described as $\sigma(\omega)=\sigma_{\rm
ni}(\omega)/(1+\sigma_{\rm ni}(\omega)/(i\omega C))$, where
$\sigma_{\rm ni}(\omega)=\sigma_\lambda/(1+i\omega\tau_\lambda)$
the non-interacting case and capacitance $C\approx e^2g$
quantifies the 1D screening efficiency. \cite{Buttiker93} For
$\hbar\omega>\Delta$ the isolated chains' response (Lorentz
oscillator) is obtained: $\sigma(\omega)=
\sigma_0/[1+i((\omega^2-\omega_0^2)/\omega)\tau_0]$ with $\omega_0$
a measure for level separation. In case the time scales are
sufficiently separated, the total optical response is
\begin{equation}
\sigma(\omega)\approx {\sigma_{\rm 3D}\over{1+i\omega\tau_{\rm 3D}}} +
{\sigma_\lambda\over{1+i\omega\tau_\lambda+{{\sigma_\lambda}\over{i\omega C}}}} +
{\sigma_0\over{1+i{{\omega^2-\omega_0^2}\over{\omega}}\tau_0}}
\end{equation}
Here $\sigma_{\rm 3D}=\varepsilon_0\omega_{\rm p}^2\tau_{\rm 3D}$,
$\sigma_\lambda=\sigma_{\rm 3D}$ because the factors $\epsilon$
in $\tau_\lambda$ and bandwidth $\Delta\exp(-\lambda/\zeta)$ cancel and
$\sigma_0=\varepsilon_0\Omega_{\rm p}^2\tau_0=ne^2\tau_0/m$ the
intrinsic intrachain conductivity.

Here $\sigma_{\rm 3D}=\varepsilon_0\omega_{\rm p}^2\tau_{\rm 3D}$,
$\sigma_\lambda=\sigma_{\rm 3D}$ because the factors $\epsilon$ in
$\tau_\lambda=\epsilon\tau_{\rm 3D}$ and $\omega_{\rm
p,\lambda}^2= \omega_{\rm p}^2/\epsilon$ cancel and
$\sigma_0=\varepsilon_0\Omega_{\rm p}^2\tau_0=ne^2\tau_0/m$ the
intrinsic intrachain conductivity.

Indeed, Fig.~\ref{optical_conductivity} demonstrates our theory
gives an excellent description of the experimental results, using
$\tau_0=3$~fs, $\tau_\lambda=0.4$~ps, $\tau_{\rm 3D}=5$~ps, $\sigma_{\rm 3D}=
\sigma_\lambda=220$~S/cm, $\sigma_0=380$~S/cm, $\omega_0=0.23$~eV, and $C=10^{-10}$~F/m.
Thus, $\lambda/\zeta=3.5$ and $\epsilon=0.08$, which assuming
$a=4$~\AA\ leads to $\zeta=1.4$~nm and $\lambda=5.0$~nm.

We conclude by addressing the scaling relation between
$\omega_{\rm p}$ and $\tau_{\rm 3D}$. Eqs. (\ref{tau3D}) and (\ref{omegap}) express
the extreme sensitivity of the 3D carrier dynamics
for disorder and interchain coupling. For increasing disorder (lower $\zeta$)
and decreasing coupling (lower $I_\perp$) the transition time to 3D conduction
increases exponentially, while simultaneously the plasma frequency decreases at the same rate.
Let us consider the case of a disorder driven MIT. We chose $\tau_0=1.2$--4.0~fs as typical
range and $\zeta=4v_{\rm F}\tau_0$,\cite{Prigodin02} to reproduce the experimental results
in Fig.~\ref{scaling} using known microscopic parameters for conducting polymers:
\cite{Handbook98} $I_\perp=50$~meV, $a=4$~\AA, $b=3$~\AA,
$g=0.5$~eV$^{-1}a^{-1}$, and $n=7\times10^{27}\rm\ m^{-3}$, which
corresponds to optimum doping of 0.25 carrier per monomer. A
similar calculation can be performed for a MIT purely driven by
interchain transfer, but agreement with experiment can be achieved
only for an unrealistically large change in $I_\perp$ of 20 to
90~meV. This agrees with the general consensus that the MIT in
polymers is disorder driven. Using the above
parameters, we have calculated the phase-diagram for the
disorder-driven MIT, see Fig.~\ref{MIT}, illustrating the
transition to the metallic state. Indeed it shows the conducting
polymers are on the edge of being metallic or insulating.

In summary, we have discussed a transport model
for weakly coupled disordered chains. Under appropriate
conditions, $(B/E_{\rm F})^2<\epsilon\ln(1/\epsilon)$, a crossover
from 1D to 3D transport occurs and a metallic phase is formed.
Expressions for the optical conductivity, 3D scattering time
and plasma frequency have been derived. Based on established microscopic
parameters, the model quantitatively explains broad-band data on
metallic PPy.

\end{document}